\documentclass[preprint,tightenlines,aps,prd,showpacs,nofootinbib]{revtex4}
\usepackage{graphicx}
\usepackage{amsmath}
\usepackage{amssymb}
\usepackage{bm}

\begin{document}

\title{\mbox{}\\[10pt]
Estimate of the Partial Width 
for $\bm{X(3872)}$ into $\bm{p \bar p}$}
%%%%%%%%%%%%%%%%%%%%%%%%%%%%%%%%%%%%%%%%%%%%%%%%%%%%%%%%%%%%%%%%%%%%%%%%%%%%%%

\author{Eric Braaten}
%\thanks{}
\affiliation{Physics Department, Ohio State University, Columbus,
Ohio 43210, USA}

\date{\today}
%%%%%%%%%%%%%%%%%%%%%%%%%%%%%%%%%%%%%%%%%%%%%%%%%%%%%%%%%%%%%%%%%%%%%%%%%%%%%%
\begin{abstract}
We present an estimate of the partial width 
of $X(3872)$ into $p \bar p$ under the assumption that  
it is a weakly-bound hadronic molecule whose constituents
are a superposition of the charm mesons
$D^{*0} \bar D^{0}$ and $D^{0} \bar D^{*0}$.
The $p \bar p$ partial width of $X$ is therefore 
related to the cross section for $p \bar p \to D^{*0} \bar D^{0}$
near the threshold. 
That cross section at an energy well above the threshold is estimated 
by scaling the measured cross section for $p \bar p \to K^{*-} K^+$.
It is extrapolated to the $D^{*0} \bar D^{0}$ threshold by taking 
into account the threshold resonance in the $1^{++}$ channel. 
The resulting prediction for the $p \bar p$ partial width of 
$X(3872)$ is proportional to the square root of its 
binding energy.  For the current central value of the 
binding energy, the estimated partial width 
into $p \bar p$ is comparable to that
of the P-wave charmonium state $\chi_{c1}$.
\end{abstract}

% 12.38.-t   Quantum chromodynamics
% 12.39.St  Factorization
% 13.20.Gd  Decays of J/psi, Upsilon, and other quarkonia
% 14.40.Gx   Mesons with S=C=B=0, mass > 2.5 GeV (including quarkonia)

%\keywords{}

\maketitle


\section{Introduction}

Since the surprising discovery of the $X(3872)$ in 2003 \cite{Choi:2003ue}, 
there has been a steadily growing list of new $c \bar c$ mesons 
discovered at the $B$ factories 
\cite{Abe:2004zs,Aubert:2005rm,Abe:2007jn,Abe:2005bp,%
Aubert:2006ge,Yuan:2007sj,Wang:2007ea,Wang:2007wga}.  
These discoveries have revealed that the 
spectrum of $c \bar c$ mesons is richer than the charmonium states 
predicted by quark potential models.  The candidates for some of the 
new $c \bar c$ mesons include charm meson molecules, 
tetraquark states, and charmonium hybrid states \cite{Swanson:2006st}.  
An ideal experiment 
for studying some of the new $c \bar c$ mesons would be 
$p \bar p$ collisions at resonance.  The effectiveness of resonant 
$p \bar p$ collisions for studying conventional charmonium states 
was demonstrated by the E760 and E835 experiments at Fermilab 
\cite{Patrignani:2004vt,Zioulas:1997ya}. 
The ``$c \bar c$ Mesons'' section of the Review of Particle Physics 
\cite{Yao:2006px} was largely rewritten by these experiments.
There will be future opportunities to study $c \bar c$ mesons 
through resonant $p \bar p$ collisions.  The Panda experiment 
at GSI is expected to begin taking data on $p \bar p$ collisions 
at energies in the charmonium region around 2014 \cite{Ritman:2007ue}.
A recently proposed $p \bar p$ resonance experiment at Fermilab 
could begin even earlier \cite{Kaplan:2007aj}.

The rate at which a resonance is produced in $p \bar p$ collisions is
proportional to its partial width into $p \bar p$.  Thus a resonant 
$p \bar p$ annihilation experiment would be useful for studying 
the new $c \bar c$ mesons provided their $p \bar p$ partial widths 
are sufficiently large.  To estimate the $p \bar p$ 
partial widths of conventional charmonium states, one can take 
advantage of the measured $p \bar p$ partial widths of the $\eta_c$, 
$J/\psi$, $\chi_{c0}$, $\chi_{c1}$, $\chi_{c2}$, and $\psi(2S)$. 
It is difficult to estimate the $p \bar p$ partial widths for most 
of the more exotic candidates for the new $c \bar c$ mesons. 
A weakly-bound S-wave charm meson molecule is an exception. 
Its $p \bar p$ partial width is related in a simple way
to the cross section near threshold for $p \bar p$ annihilation 
into the charm mesons that are its constituents.
The energy dependence of that cross section near threshold is 
determined by the mass and width of the resonance. 
If the S-wave contribution to the cross section for $p \bar p$ 
annihilation into the charm mesons 
well above the threshold was known, 
a reasonable extrapolation to the threshold could be made 
in terms of the position and width of the resonance. 
Cross sections for $p \bar p$ annihilation into pairs of
charm mesons have not been measured.  However at energies well above 
threshold, those cross sections 
can be estimated by scaling the measured cross sections for $p \bar p$
annihilation into the corresponding strange mesons.
In this paper, we will use this strategy to estimate the
partial width of the $X(3872)$ into $p \bar p$.

We will asume in this paper that the $X(3872)$ is a weakly-bound
hadronic molecule whose constituents are a superposition of 
$D^{*0} \bar D^0$ and $D^0 \bar D^{*0}$. 
We first summarize the evidence for this identification. 
Measurements of the 
mass of the $X(3872)$ by the Belle, CDF, Babar, and D0 collaborations
\cite{Choi:2003ue,Acosta:2003zx,Abazov:2004kp,Aubert:2004ns},
combined with a new measurement of the $D^0$ mass by the CLEO 
collaboration \cite{Cawlfield:2007dw}, imply that the mass is extremely 
close to the $D^{*0} \bar D^0$ threshold:
\begin{equation}
M_X - (m_{D^*} + m_{\bar D^0}) = -0.6 \pm 0.6 ~ \rm{MeV} .
\label{MX}
\end{equation}
The observation of the decay of $X$ into $J/\psi \, \gamma$  
\cite{Abe:2005ix} implies that the charge conjugation 
quantum number is $C=+1$. Studies of the decays of $X$ into 
$J/\psi \, \pi^+ \pi^-$ \cite{Abe:2005iy,Abulencia:2005zc} 
strongly favor the spin and parity $J^P = 1^+$, 
although $2^-$ is not excluded.  The observation of decays into
$D^0 \bar D^0 \pi^0$ \cite{Gokhroo:2006bt,Aubert:2007rv} 
excludes $J \ge 2$, because the available phase space is so small. 
Thus the quantum numbers of the $X(3872)$ have been determined to be 
$J^{PC} = 1^{++}$. These quantum numbers imply that the $X$ has an
S-wave coupling to the charm meson channel 
$D^{*0} \bar D^0 + D^0 \bar D^{*0}$.   The mass measurement 
in Eq.~(\ref{MX}) implies that it is a resonant coupling.
This is sufficient to conclude 
that the $X(3872)$ is either a charm meson molecule with particle 
content $D^{*0} \bar D^0 + D^0 \bar D^{*0}$ 
or it is a virtual state of the charm mesons.
The reason such an unambiguous statement can be made is that
nonrelativistic systems with S-wave threshold resonances have universal
properties that depend on the large scattering length of the constituents
but are otherwise insensitive to their interactions at shorter 
distances \cite{Braaten:2004rn}. These universal properties 
have been exploited to describe the decays of $X$ into 
$D^0 \bar D^0 \pi^0$ and
$D^0 \bar D^0 \gamma$ \cite{Voloshin:2003nt}, the
production process $B \to KX$ \cite{Braaten:2004fk,Braaten:2004ai},
the line shapes of the $X$ \cite{Braaten:2005jj}, and
decays of $X$ into $J/\psi$ and pions \cite{Braaten:2005ai}.
These applications rely on factorization formulas that separate
the length scale $a$ from all the shorter distance scales of QCD
\cite{Braaten:2005jj,Braaten:2006sy}.
Hanhart et al.~have analyzed the data from the Belle and Babar
collaborations on the decays $B^+ \to K^+ + X$ in the decay channels
$J/\psi \, \pi^+ \pi^-$ and $D^0 \bar D^0 \pi^0$ and concluded that the $X$
must be a virtual state of charm meson \cite{Hanhart:2007yq}. 
A more recent analysis that took into account consistently the effects 
of the $D^{*0}$ width concluded that a charm meson molecule 
was preferred by the data, although a virtual state
could not be excluded \cite{Braaten:2007dw}.

In this paper, we exploit the identification of the $X(3872)$ 
as a weakly-bound charm meson molecule to give an order-of-magnitude 
estimate of its partial width into $p \bar p$. 
In Section~\ref{sec:hi-energy}, we estimate the cross
section for $p \bar p \to D^{*0} \bar D^0$ well above the threshold by
scaling measured cross sections for $p \bar p \to K^{*-} K^+$. 
In Section~\ref{sec:threshold}, we
extrapolate the cross sections for $p \bar p \to D^{*0} \bar D^0$ to the
threshold under the assumption that there is an S-wave threshold
resonance in the $1^{++}$ channel. In Section~\ref{sec:width}, we deduce 
the partial width for $X$ into $p \bar p$ from the extrapolated 
cross section for $p \bar p \to D^{*0} \bar D$. 
We discuss the results in Section~\ref{sec:disc}.

\section{Cross Section for $\bm{p \bar p \to D^{*0} {\bar D}^0}$ 
	Far Above Threshold}
\label{sec:hi-energy}

Given that $X(3872)$ is a weakly-bound S-wave hadronic molecule 
composed of $D^{*0} \bar D^{0}$ or $D^{0} \bar D^{*0}$, 
its partial width into $p \bar p$ is proportional to the cross section
for $p \bar p \to D^{*0} \bar D^{0}$ near the $D^{*0} \bar D^{0}$ threshold. 
In this section, we estimate that cross section at energies 
well above the threshold by scaling the measured cross section 
for $p \bar p \to K^{*-} K^+$.

At the quark level, the process $p \bar p \to D^{*0} \bar D^0$ is 
$uud +\bar u \bar u \bar d \to c \bar u + \bar c u$. 
This process can proceed through
the short-distance annihilation process $ud + \bar u \bar d \to c + \bar c$
followed by a long-distance process in which the remaining constituents
$\bar u$ and $u$ of the $\bar p$ and $p$ bind to the $c$ and $\bar c$ to
form the charm mesons. The $\bar u$ and $u$ carry only a
small fraction of the momenta of the outgoing charm mesons. 
This process can proceed most easily via the Feynman process in which
the $u$ and $\bar u$ that do not annihilate carry only a small 
fraction of the momenta of the colliding $p$ and $\bar p$.
The momenta transfered to the $u$ and to the $\bar u$ are therefore small.
We assume that the cross section for this process can be factored 
into the rate for the short-distance annihilation process 
$ud + \bar u \bar d \to c + \bar c$ and a long-distance probability factor
associated with the $u$ and $\bar u$ that do not annihilate.

We first consider the short-distance factor.  The
short-distance process involves the annihilation of $ud + \bar u \bar d$
into two virtual gluons which then create a $c \bar c$ pair. Since it
involves 6 external partons, the dimensional counting rules for this process
imply that its rate scales with the center-of-mass energy $\sqrt{s}$ like
$1/s^3$ \cite{Brodsky:1973kr,MMT:1973}. 
This should be contrasted with the rate for $p \bar p$ annihilation
at resonance into charmonium. In this case, the short-distance process is
$uud + \bar u \bar u \bar d \to c \bar c$.  
The dimensional counting rules imply
that its rate scales like $1/M^8$, 
where $M$ is the charmonium mass \cite{Brodsky:1981kj}. 
In the case of $p \bar p \to D^{*0} \bar D^0$,
the rate for the short-distance subprocess scales
with two fewer powers of the center-of-mass energy, because one of the
three quarks in the proton is not required to annihilate.

We now consider the long-distance factor.
This factor is the probability for the 
surviving $u$ and $\bar u$ from the colliding $p$ and $\bar p$
to become constituents of the outgoing charm mesons. 
In the initial state, the $ud$ and the $\bar u \bar d$ that 
annihilate act as light-like colored sources
for the remaining low-momentum $u$ and $\bar u$. In the final state, 
the $c$ and $\bar c$ act as colored sources for the
low-momentum $\bar u$ and $u$.
In the $p \bar p$ rest frame, these sources have equal and opposite
velocities $\beta = [(s - 4m_c^2)/s]^{1/2}$.  
The effect of the short-distance
process is to suddenly replace the light-like colored sources  
$ud$ and $\bar u \bar d$ by colored sources $c$ and $\bar c$ with equal 
and opposite velocities $\beta$.  The long-distance factor $P(\beta)$
is the square of the amplitude for the low-momentum $u$ and $\bar u$
to evolve from constituents of the $p$ and $\bar p$ into constituents 
of the $\bar D^0$ and $D^{*0}$ after the sudden change in color sources.

The long-distance factor $P(\beta)$ depends on the velocity $\beta$ 
of the colored sources in the final state, 
but it does not depend on the flavor of the particles
that serve as the colored sources. The amplitude would be the same if the
charm quarks $c$ and $\bar c$ were replaced by strange quarks $s$ and
$\bar s$ with the same velocities $\beta$. Thus the rate for producing the
charm mesons $D^{*0}+\bar D^0$, whose quark content is 
$c \bar u + \bar c u$,  can be related to the rate for producing
strange mesons $K^{*-}+K^+$, whose quark content is $s \bar u + \bar s u$.
If we take into account the masses of the mesons, a $K^{*-} K^+$ pair with
center-of-mass energy $s_K^{1/2}$ has the same relative velocity 
$2 \beta$ in the
center-of-mass frame as a $D^{*0} \bar D^0$ pair with center-of-mass
energy $s_D^{1/2}$ if the center-of-mass energies satisfy
\begin{equation}
\frac {s_K ~ \lambda^{1/2}(s_K^{1/2}, m_{K^*}, m_K)}
{s^2_K - (m_{K^*}^2 - m_K^2)^2} =
\beta =  
\frac {s_D ~ \lambda^{1/2}(s_D^{1/2}, m_{D^*}, m_D)}
{s^2_D - (m_{D^*}^2 - m_D^2)^2} ,
\label{sD-sK}
\end{equation}
% y^2
where $\lambda(x,y,z) = x^4 + y^4 + z^4 -2(x^2 y^2 + y^2 z^2+ z^2 x^2)$.

The cross section for $p \bar p \to D^{*0} \bar D^0$
can be expressed as the product of a flux factor 
$[s (s - 4 m_p^2)]^{-1/2}$, a phase space factor 
$\lambda^{1/2}(s^{1/2}, m_{D^*}, m_D) /(8 \pi s)$,
and a matrix element factor.  Under our factorization assumption, 
the matrix element factor is the product of a short-distance 
factor that scales asymptotically like $s^{-2}$  
and a long-distance factor $P(\beta)$ that is a function of the 
velocities $\beta$  of the $c$ and $\bar c$ created by the 
short-distance process.  If we use the asymptotic scaling behavior 
of the short-distance factor in the matrix element, the energy
dependence of the cross section is given by
\begin{eqnarray}
\sigma [ p \bar p \to D^{*0} \bar D^0; s ] &\sim& 
\frac{1}{[s (s - 4 m_p^2)]^{1/2}} \,
\frac{P(\beta)}{s^2}  \,
\frac{\lambda^{1/2}(s^{1/2}, m_{D^*}, m_D)}{s} .
\label{sig-E}
\end{eqnarray} 
In the limit $s \to \infty$, the right side of Eq.~(\ref{sig-E})
approaches $P(1)/s^3$ in accord with the dimensional counting rules.
There is an expression analogous to  Eq.~(\ref{sig-E})
for the cross section for 
$p \bar p \to K^{*-} K^+$ with the same long-distance factor
$P(\beta)$, but with $m_K$ and $m_{K^*}$ replaced by $m_D$ and $m_{D^*}$.  
Using Eq.~(\ref{sD-sK}) to eliminated that factor, 
we obtain a scaling relation between the two cross sections:
\begin{eqnarray}
\sigma [ p \bar p \to D^{*0} \bar D^0; s_D ] &\approx& 
\sigma [ p \bar p \to K^{*-} K^+ ; s_K ] \,
\frac {[s_K (s_K - 4 m_p^2)]^{1/2}}  
     {[s_D (s_D - 4 m_p^2)]^{1/2}}
\nonumber
\\
&& \hspace{1cm}
\times \left( \frac {s_K}{s_D} \right)^3 
\frac {\lambda^{1/2}(s_D^{1/2}, m_{D^*}, m_D)}
     {\lambda^{1/2}(s_K^{1/2}, m_{K^*}, m_K)} ,
\label{sig-scale}
\end{eqnarray} 
where $s_K$ is the function of $s_D$ obtained by solving
Eq.~(\ref{sD-sK}).

%------------------------------------------------------------------------------------
\begin{figure}[t]
\includegraphics[width=12cm]{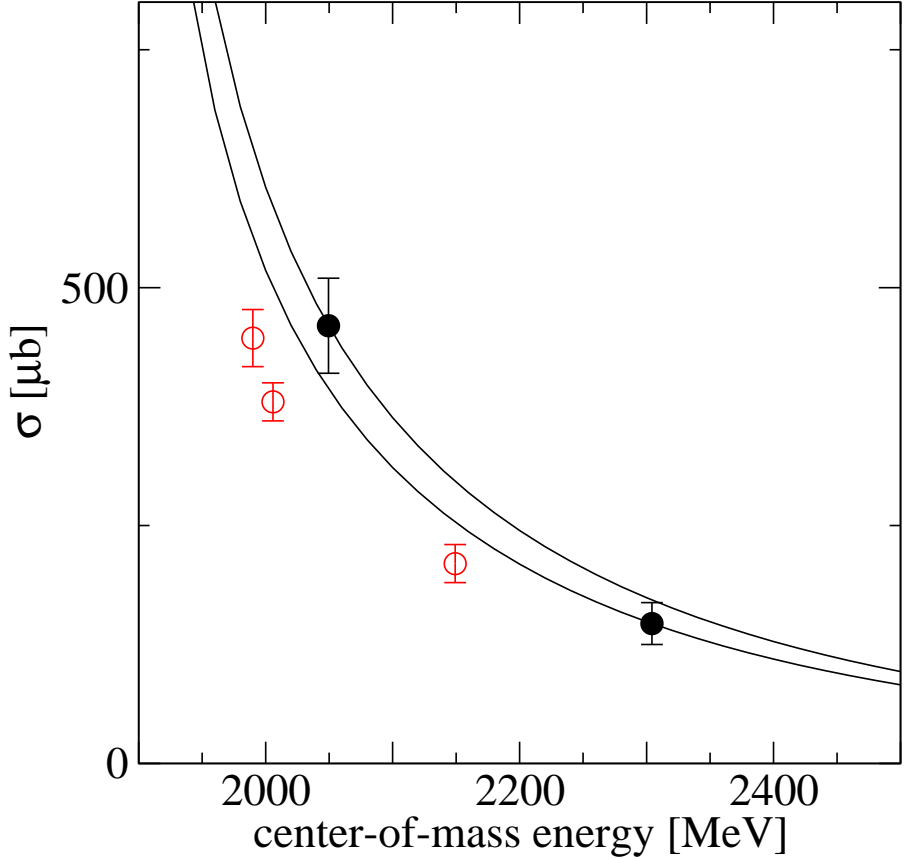}
\caption{
Measured cross sections for $p \bar p \to K^{*-} K^+$ as a 
function of the center-of-mass energy.  The open dots are
the data from Refs.~\cite{barlow,Ganguli:1980vr}.  The two solid dots 
are data from the Crystal Barrel Collaboration 
\cite{Amsler:2006du}.  The curves passing through the 
Crystal Barrel data points are extrapolations whose energy 
dependence is given by the analog of Eq.~(\ref{sig-E}) 
for $K^{*-} K^+$ with the probability factor $P(\beta)$ 
approximated by a constant.
\label{fig:sigmaK}}
\end{figure}
%------------------------------------------------------------------------------------

The cross sections for $p \bar p \to K^{*-} K^+$ have been measured for
antiproton momenta in the lab frame ranging from 702 MeV 
to 1642 MeV \cite{barlow,Ganguli:1980vr,Amsler:2006du}. 
This corresponds to center-of-mass energies $\sqrt s$
ranging from 1990 MeV to 2304 MeV. 
The only recent measurements were from the 
Crystal Barrel Collaboration \cite{Amsler:2006du}. 
The cross section was measured to be $(460 \pm 50)~\mu$b
at $\sqrt s = 2050$ MeV and $(147 \pm 22)~\mu$b
at 2304 MeV \cite{Amsler:2006du}.  The relative velocities $\beta$ 
given by Eq.~(\ref{sD-sK}) are 0.74 and 0.80, respectively.
The data points are shown in
Figure~\ref{fig:sigmaK} as a function of the center-of-mass energy. 
Also shown in the figure are curves that pass through the Crystal Barrel
data points and whose dependence on the energy is given by 
the analog of Eq.~(\ref{sig-E}) for $K^{*-} K^+$ with the 
probability factor $P(\beta)$ approximated by a constant.
Notice that the two curves do not differ dramatically. 
This suggests that at these energies, 
the matrix element factor in the $K^{*-} K^+$ cross section
may already be close to its asymptotic behavior 
proportional to $P(1)/s^3$. 

To estimate the cross section for $p \bar p \to D^{*0} \bar D^0$,
we insert the Crystal Barrel data points for the $K^{*-} K^+$ 
cross section into Eq.~(\ref{sig-scale}), where $s_K$ is the
function of $s = s_D$ obtained by solving Eq.~(\ref{sD-sK}).
The center-of-mass energies $s_K^{1/2}$ for the two Crystal Barrel 
data points are 2050 MeV and 2304 MeV.  The corresponding 
center-of-mass energies $s_D^{1/2}$ for the process
$p \bar p \to D^{*0} \bar D^0$ are 5714 MeV and 6391 MeV.
If we use the higher-energy Crystal Barrel data point
as the input, our estimate for the cross section 
for $D^{*0} \bar D^0$ is
\begin{equation}
\sigma [ p \bar p \to D^{*0} \bar D^0; s] \approx 
\left( \frac{m_{D^*}+m_D}{\sqrt{s}} \right)^6 \,
\frac{\lambda^{1/2}(s^{1/2}, M_{D^*}, M_D)}
    {[s (s - 4 m_p^2)]^{1/2}} 
\left( 4800 ~ {\rm nb} \right)  .
\label{sig-hi}
\end{equation}
If we use the lower-energy Crystal Barrel data point as the input,
the factor of 4800 nb is replaced by 5600 nb.  
The relatively small difference implies that the 
probability factor $P(\beta)$ in Eq.~(\ref{sig-E}) does not depend 
dramatically on $\beta$ in the region $0.74 < \beta < 0.80$.
If we were to change the 
assumed scaling behavior of the short-distance factor 
in the square of the matrix element from
$(\sqrt{s})^{-6}$ to $(\sqrt{s})^{-6 \pm 1}$, the prediction 
in Eq.~(\ref{sig-hi}) would change by a multiplicative factor of 
$2.8^{\pm 1}$.

\section{Cross Section for $\bm{p \bar p \to D^{*0} {\bar D}^0}$ 
        near Threshold}
\label{sec:threshold}

Given the estimate in Eq.~(\ref{sig-hi}) of the cross section for
$p \bar p \to D^{*0} \bar D^0$ at energies far above threshold,
we would like to obtain an estimate of the cross section near 
threshold.  We cannot simply use Eq.~(\ref{sig-hi}) to extrapolate 
to the threshold region, 
because this expression ignores the suppression of higher partial waves 
near the threshold, which can be taken into account through the probability 
factor $P(\beta)$ in Eq.~(\ref{sig-E}), and because it also ignores 
the threshold resonance associated with the $X(3872)$.
We will assume that the threshold resonance is in the 
S-wave $C=+1$ channel, and that this is the only effect that gives 
dramatic dependence of the S-wave phase shift on the
center-of-mass energy $\sqrt s$.

As the center-of-mass energy approaches the threshold at 
$\sqrt s = M_{D^{*0}} + M_{D^0}$, the terms in the matrix element 
with nonzero orbital angular momentum quantum number $L$ 
are suppressed by a factor of $(\sqrt s - M_{D^{*0}} - M_{D^0})^{L/2}$. 
Only the S-wave $L=0$ term survives in the limit. 
If the S-wave threshold resonance could be ignored, 
a crude extrapolation of the cross section for 
$D^{*0} \bar D^0$ in Eq.~(\ref{sig-hi}) towards the threshold 
could be obtained simply by multiplying it by the fraction 
$f_{L = 0}$ of the cross section that comes from S-wave scattering.
The determination of the S-wave fraction requires measurements 
of the angular distribution.  There is some information on the 
angular distribution for the related process $p \bar p \to K^{*-} K^+$.
In the Crystal Barrel experiment \cite{Amsler:2006du}, 
it was estimated that angular momenta up to $L=5$ contribute 
to the cross section at center-of-mass energy 2304 MeV.  
In Ref.~\cite{Ganguli:1980vr}, the angular distribution for 
$p \bar p \to K^{*-} K^+$ was measured at center-of-mass energy 2006 MeV.
The ratios $a_L/a_0$ of the coefficients of the
Legendre polynomials in the partial wave expansion
were given for $L$ up to 5. From those results, 
we can infer that the fraction of the cross section at 2006 MeV that
comes from S-wave scattering is about 9.3\%. 
Since no experimental information is available on the S-wave fraction 
$f_{L=0}$ for $D^{*0} \bar D^0$, we will use the S-wave fraction
for $K^{*-} K^+$ at 2006 MeV as an estimate: 
$f_{L=0} \approx 9.3\%$.

If the threshold resonance could be ignored, we could estimate 
the cross section for $D^{*0} \bar D^0$ near threshold
obtained simply by extrapolating Eq.~(\ref{sig-hi}) to the threshold 
and multiplying it by the fraction $f_{L = 0}$.
However the existence of the $X(3872)$ with quantum numbers $1^{++}$ 
implies that there is a resonant enhancement of the cross section 
for $D^{*0} \bar D^0$ near the threshold due to scattering in the 
S-wave $C=+1$ channel.
The enhancement applies only to the fraction $f_{C=+1}$ of the cross section 
that comes from $p \bar p$ scattering into channels 
that are even under charge conjugation. 
Experiments on $p \bar p$ annihilation at very low $\bar p$ momentum have
revealed a suppression of the production of final states that are
even under charge conjugation.
In Ref.~\cite{Ganguli:1980vr}, the suppression of
$C = +1$ final states was also observed in final states containing pairs of 
strange mesons at the center-of-mass energy 2006 MeV. 
The suppression factor in the case of
$K^{*-} K^+$ can be deduced from measurements of $K^{*0} K^0$, which is
related by isospin symmetry. The cross section for $p \bar p \to K^{*0}
K_S^0$ followed by the decay $K^{*0} \to K_S^0 \pi^0$, which gives a final
state with $C = +1$, is only $17 \pm 4$ $\mu$b.
The cross section for $p \bar p \to K^{*0} K_S^0$ with no restriction on the
decays of $K^{*0}$ is $130 \pm 10$ $\mu$b.  The ratio of these two
cross sections gives an even-charge-conjugation fraction 
of about 13.1\%. 
Since no experimental information is available on the 
even-charge-conjugation fraction $f_{C = +1}$ for $D^{*0} \bar D^0$, 
we will use the even-charge-conjugation fraction
for $K^{*0} K_S^0$ at 2006 MeV as an estimate: 
$f_{C = +1} \approx 13.1\%$.

To extrapolate the cross section for $p \bar p \to D^{*0} \bar D^0$
to the threshold region, we will exploit the universal features 
of S-wave threshold resonances in nonrelativistic 2-body systems
\cite{Braaten:2004rn}.
These universal features can be conveniently expressed in
terms of the scattering length $a$ in the resonant S-wave channel.
If the resonance is sufficiently close to the threshold, 
$|a|$ will be much larger than all other length scales associated 
with the structure of the particles and their interactions.
The scattering  amplitude in the resonant channel has the universal form
\begin{equation}
f(E) = \frac{1}{-1/a + \sqrt{- 2 M E - i \epsilon}}, 
\label{f-E}
\end{equation}
where $E$ is the energy relative to the 2-body threshold
and $M$ is the reduced mass.  If $a$ is real and positive, 
the amplitude in Eq.~(\ref{f-E}) has a pole at $E = -E_X$, where
\begin{equation}
E_X = \frac {1}{2M a^2} .
\label{EX}
\end{equation}
The pole is associated with a bound state with binding energy $E_X$.
If $a$ is real and negative, there is no pole on the physical sheet of the
complex energy $E$, so there is no bound state.  There is however 
a pole on the unphysical sheet of $E$ and it is conventionally 
referred to as a ``virtual state''.  The scattering length $a$
is complex if the 2-body system has inelastic scattering channels.  
If $a$ is complex, its real part can still be used to distinguish 
between a bound state (${\rm Re} \, a > 0$)
and a virtual state (${\rm Re} \, a < 0$).
In the case ${\rm Re} \, a > 0$, the bound state has a nonzero width 
determined by the imaginary part of $a$.

The existence of the $X(3872)$ with quantum numbers $1^{++}$ 
and mass given by Eq.~(\ref{MX}) implies that there is an S-wave 
threshold resonance in the channel
\begin{equation}
(D^* \bar D)_+^0 = 
\frac{1}{\sqrt{2}} \left( D^{*0} \bar D^0 + D^{0} \bar D^{*0} \right).
\label{DstarDbar}
\end{equation}
The scattering amplitude in this channel is given by the universal 
expression in Eq.~(\ref{f-E}), where 
$E = \sqrt{s} - m_{D^{*0}} - m_{D^0}$ and $M$ is the reduced mass
$M_{D^* \bar D} = m_{D^{*0}} m_{D^0} / (m_{D^{*0}} + m_{D^0})$.
Since $X(3872)$ decays, its constituents have inelastic scattering 
channels, so the scattering length $a$ must be complex.  
In Ref.~\cite{Hanhart:2007yq}, the authors analyzed the data 
from the Belle and Babar collaborations on the $X(3872)$ resonance 
produced via $B^+ \to K^+ + X$ and decaying through
the channels $J/\psi \, \pi^+ \pi^-$ and $D^0 \bar D^0 \pi^0$.
They concluded that $X(3872)$ must be a virtual state with 
${\rm Re} \, a < 0$.  In a subsequent analysis of the same data,
the nonzero width of the constituent $D^{*0}$ or $\bar D^{*0}$
of the $X$ was taken into account \cite{Braaten:2007dw}.  The 
conclusion of that analysis was that the data preferred a 
bound state corresponding to ${\rm Re} \, a > 0$, 
although a virtual state was not excluded.  
We will assume that the $X(3872)$ is indeed a bound state.
We also assume for simplicity that the imaginary part of $a$ 
is small compared to its real part, so the binding energy of the $X$
can be approximated by the simple expression in Eq.~(\ref{EX}) 
with $M$ replaced by $M_{D^* \bar D}$.

We proceed to use the universality of S-wave threshold resonances 
to extrapolate the estimated cross section for 
$p \bar p \to D^{*0} \bar D^0$ to the threshold region.
Matrix elements corresponding to the projection of the 
charm mesons onto the channel 
$(D^* \bar D)_+^0$ defined in Eq.~(\ref{DstarDbar}) 
will be enhanced at energies near the $D^{*0} \bar D^0$ threshold
by the resonance associated with the $X(3872)$.  For the process
$p \bar p \to D^{*0} \bar D^0$, the resonance enhances 
the matrix element near the threshold by a dimensionless factor
\begin{equation}
(2/\pi)\Lambda f(E) = \frac{(2/\pi)\Lambda}{-1/a - i k_{\rm cm}}, 
\end{equation}
where
$k_{\rm cm} = (2 M_{D^{*} \bar D} E)^{1/2}$ and $\Lambda$ 
is the momentum scale above which the effects of the
nonzero range of the interaction become important.
If the S-wave effective range $r_s$ for the scattering of the 
charm mesons was known, we could use $\Lambda \approx 2/r_s$ 
as a quantitative estimate.
A natural order-of-magnitude estimate for $\Lambda$ is $m_\pi$, 
which is the next smallest relevant momentum
scale after $1/|a|$.  This is a relevant momentum scale, 
because the low energy scattering of $D^{*0} \bar D^0$
is dominated by pion exchange.
Our estimate of the cross section for 
$p \bar p \to D^{*0} \bar D^0$ near the threshold is obtained 
by extrapolating the estimated cross section in Eq.~(\ref{sig-hi})
to the threshold region and multiplying it by  
the S-wave fraction $f_{L=0} \approx 9.3\%$,
the even-charge-conjugation fraction $f_{C = +1} \approx 13.1\%$, 
and the resonance factor $(4/\pi^2) \Lambda^2 |f(E)|^2$:
\begin{equation}
\sigma [ p \bar p \to D^{*0} \bar D^0; E ] \approx
\left( 1.9 ~ {\rm nb} \right) 
\frac{\Lambda^2 k_{\rm cm}}{m_\pi (1/a^2 + k_{\rm cm}^2)}  .
\label{sigD*Dbar0}
\end{equation}
The factor of $m_\pi$ in the denominator in Eq.~(\ref{sigD*Dbar0})
was introduced to make the last factor dimensionless.
If the lower-energy Crystal Barrel data point was used as the 
input for Eq.~(\ref{sig-hi}), the prefactor in Eq.~(\ref{sigD*Dbar0})
would have been 5.5 nb.

\section{Partial width into $\bm{p \bar p}$}
\label{sec:width}

To relate the cross sections for $p \bar p$ annihilation into 
$D^{*0} \bar D^0$ and $X(3872)$, we begin by considering the 
forward scattering amplitude for $p \bar p \to p \bar p$
near the $D^{*0} \bar D^0$ threshold.  This amplitude can be 
expressed as the product of three factors:
a short-distance factor for the transition from $p \bar p$
to the resonant channel $(D^* \bar D)_+^0$, 
a long-distance resonance factor $f(E)$ given by Eq.~(\ref{f-E}),
and a short-distance factor for the transition from 
$(D^* \bar D)_+^0$ to  $p \bar p$.  By the optical theorem, 
the total $p \bar p$ cross section is proportional to the 
imaginary part of the forward scattering amplitude.
The resonant contribution to the imaginary part associated with 
$D^{*0} \bar D^0$ or $D^0 \bar D^{*0}$ in the final state
comes from the imaginary part of the resonance factor:
\begin{equation}
{\rm Im} f(E) =
\frac{k_{\rm cm}}{1/a^2 + k_{\rm cm}^2} \theta(E)
+ \frac{\pi}{M_{D^* \bar D} a} \delta(E + E_X) ,
\end{equation}
where 
$k_{\rm cm} = (2 M_{D^{*}\bar D} E)^{1/2}$
and $E_X = 1/(2M_{D^* \bar D} a^2)$ is the binding energy of the $X$
given in Eq.~(\ref{EX}).
The first term on the right side corresponds to the production of
$D^{*0} \bar D^0$ and $D^0 \bar D^{*0}$ above the threshold.  
The second term corresponds to the production of the $X(3872)$ resonance.
The cross sections for producing $D^{*0} \bar D^0$ near threshold,
$D^0 \bar D^{*0}$ near threshold, and $X(3872)$ are
\begin{subequations}
\begin{eqnarray}
\sigma[ p \bar p \to D^{*0} \bar D^0;E] &=& 
\mbox{$\frac12$} \sigma_{\rm SD} 
\frac{m_\pi k_{\rm cm}}{1/a^2 + k_{\rm cm}^2} \theta(E) ,
\label{sigD*Dbar-fact}
\\
\sigma[ p \bar p \to D^0 \bar D^{*0};E] &=& 
\mbox{$\frac12$} \sigma_{\rm SD} 
\frac{m_\pi k_{\rm cm}}{1/a^2 + k_{\rm cm}^2} \theta(E) ,
\label{sigDDbar*-fact}
\\
\sigma[ p \bar p \to X;E] &=& 
\sigma_{\rm SD} 
\frac{\pi m_\pi (2 M_{D^* \bar D} E_X)^{1/2}}{M_{D^* \bar D}} 
\delta(E + E_X) ,
\label{sigXfact}
\end{eqnarray}
\end{subequations}
where $\sigma_{\rm SD}$ is a short-distance factor.
In Eq.~(\ref{sigXfact}), we have used Eq.~(\ref{EX}) to eliminate
the scattering length $a$ in favor of the binding energy $E_X$. 
A factor of $m_\pi$ has been inserted into the long distance factor 
to ensure that $\sigma_{\rm SD}$ has the dimensions 
of a cross section.
The delta function of the energy in Eq.~(\ref{sigXfact}) arises 
from neglecting the width of the $X(3872)$ resonance.
An estimate for the short-distance factor $\sigma_{\rm SD}$
can be obtained
by comparing the cross section in Eq.~(\ref{sigD*Dbar-fact})
with the estimated cross section in Eq.~(\ref{sigD*Dbar0}):
\begin{equation}
\sigma_{\rm SD} = 
\frac{2 \Lambda^2}{m_\pi^2} \left( 1.9 \ {\rm nb} \right) .
\label{sigma-SD}
\end{equation}

The cross section for $p \bar p \to X$ in the narrow resonance limit 
is related to the partial width for $X \to p \bar p$ 
by simple kinematics.  If we denote the T-matrix element for
$X \to p \bar p$ by ${\cal M}$, the partial width into $p \bar p$ is
\begin{equation}
\Gamma[X \to p \bar p] = 
\frac{1}{2 M_X} \, \frac{1}{3} \sum_{\rm spins} | {\cal M} |^2 \,
\frac{(M_X^2 - 4 m_p^2)^{1/2}}{8 \pi M_X} .
\label{Gamma-T}
\end{equation}
The cross section for $p \bar p \to X$ in the narrow resonance limit
averaged over the spins of the proton and antiproton is 
\begin{equation}
\sigma[ p \bar p \to X;E] = 
\frac{1}{2 M_X (M_X^2 - 4 m_p^2)^{1/2}}  \, 
\frac{1}{4} \sum_{\rm spins} | {\cal M} |^2 \,
\frac{\pi}{M_X} \delta(E + E_X) .
\label{sigma-T}
\end{equation}
Eliminating the T-matrix element from Eqs.~(\ref{Gamma-T}) 
and (\ref{sigma-T}), we obtain
\begin{equation}
\sigma[ p \bar p \to X;E] = 
\frac{6 \pi^2 \Gamma[X \to p \bar p]}{M_X^2 - 4 m_p^2} 
\delta(E + E_X) .
\label{sigma-Gamma}
\end{equation}
Comparing with Eq.~(\ref{sigXfact}) and using the estimate 
of the short-distance factor $\sigma_{\rm SD}$ in Eq.~(\ref{sigma-SD}),
we obtain an estimate of the partial width of $X$ into $p \bar p$: 
\begin{equation}
\Gamma[X \to p \bar p] =
\frac{(M_X^2 - 4 m_p^2) \Lambda^2 (2 M_{D^* \bar D} E_X)^{1/2}}
    {3 \pi m_\pi M_{D^{*} \bar D}}
\left( 1.9 \ {\rm nb} \right) .
\label{GamXppbar}
\end{equation}
The partial width is proportional to the square root of the binding energy 
$E_X$.  The current measurement of the binding energy is 
given by Eq.~(\ref{MX}):  $E_X = 0.6 \pm 0.6$ MeV.
The expression in Eq.~(\ref{GamXppbar}) can be reduced to
\begin{equation}
\Gamma[X \to p \bar p] =
\left( \frac{\Lambda}{m_\pi} \right)^2
\left( \frac{E_X}{0.6 \ {\rm MeV}} \right)^{1/2}
\left( 28  \ {\rm eV} \right)  .
\label{GamXppbar:est}
\end{equation}
The total width of the $X$ must be greater than the width of 
its constituent $D^{*0}$, which is about 70 keV.
Thus the estimate in Eq.~(\ref {GamXppbar:est})
implies that the branching fraction for $X \to p \bar p$ 
is less than about $4 \times 10^{-4}$.

\section{Discussion}
\label{sec:disc}

We have estimated the partial width of $X(3872)$ into $p \bar p$. 
The estimate was obtained by a sequence of well-motivated steps. 
The first step was estimating the cross section for 
$p \bar p \to D^{*0} \bar D^0$ well above the threshold 
by scaling measured cross sections for $p \bar p \to K^{*-} K^+$. 
The second step was estimating the contribution from the
S-wave $1^{++}$ channel using suppression factors $f_{L=0}$ 
and $f_{C = +1}$ for the $K^{*-} K^+$ process. 
The third step was extrapolating the cross section for 
$p \bar p \to D^{*0} \bar D^0$ to the $D^{*0} \bar D^0$
threshold under the assumption that there is an S-wave threshold
resonance in the $1^{++}$ channel. The cross section for 
$p \bar p \to X$ was obtained from the cross section for 
$p \bar p \to D^{*0} \bar D^0$ near threshold by using the 
universal properties of an S-wave threshold resonance. 
The partial width for $X \to p \bar p$ is related to the
cross section for $p \bar p \to X$ by simple kinematic factors. 
Our estimate for the partial width is given in Eq.~(\ref{GamXppbar:est}).

It is widely believed that the production of a loosely-bound 
molecular state in a process involving energies much greater 
than the binding energy should be strongly suppressed. 
There is some suppression, but it is not as strong as one might 
expect. In the estimate in Eq.~(\ref{GamXppbar:est}), the
suppression is taken into account by the factor proportional to
$E_X^{1/2}$. This factor goes to $0$ as $E_X \to 0$, suggesting that
there would be complete suppression in this limit. However the
suppression factor has this simple form proportional to $E_X^{1/2}$ 
only if $E_X$ is greater than $\Gamma_X/2$, where $\Gamma_X$ 
is the total width of the $X$.  There is no further suppression 
once $E_X$ becomes comparable to $\Gamma_X/2$. A lower bound on 
$\Gamma_X$ is provided by the width of the constituent $D^{*0}$ 
or $\bar D^{*0}$: $\Gamma [D^{*0}] = 66 \pm 15$ keV. 
Thus even if the binding energy is much smaller than 0.6 MeV, 
the associated suppression factor cannot decrease the estimate in
Eq.~(\ref{GamXppbar:est}) by more than about a factor of 0.2.

The estimate for the partial width for $X(3872) \to p \bar p$ in
Eq.~(\ref{GamXppbar:est}) is proportional to $\Lambda^2$, where $\Lambda$ 
is the momentum scale above which the effects of the nonzero range 
of the interaction between the charm mesons becomes important. 
Thus the rate is very sensitive to $\Lambda$. The scale $\Lambda$ 
can be identified with $2/r_s$, where $r_s$ is the effective range 
for the scattering of the charm mesons, 
which is not known. We have suggested that a 
natural scale for $\Lambda$ is $m_\pi$, since low-energy scattering 
of charm mesons is dominated by pion exchange. 
The estimates of the rate for $B \to K + X$ in
Refs.~\cite{Braaten:2004fk,Braaten:2004ai} 
were also proportional to $\Lambda^2$. 
In this case, the choice $\Lambda = m_\pi$ may underestimate 
the rate by about an order of magnitude. Thus setting 
$\Lambda = m_\pi$ in Eq.~(\ref{GamXppbar:est}) 
may also underestimate the partial 
width of $X \to p \bar p$ by an order of magnitude.

Since the P-wave charmonium state $\chi_{c1}$ has the same quantum 
numbers $1^{++}$ as the $X(3872)$, 
it is useful to compare the $p \bar p$ partial widths
of the $X(3872)$ with that of $\chi_{c1}$,
which is $60\pm 5$~eV. 
If we set $\Lambda = m_\pi$ and $E_X = 0.6$ MeV in 
Eq.~(\ref{GamXppbar:est}), our estimate of the $p \bar p$ 
partial width of $X(3872)$ is within a factor of 2 of that of $\chi_{c1}$. 
The E760 experiment on charmonium production at resonance 
in $p \bar p$ collisions measured the total width of the 
$\chi_{c1}$ with an uncertainty of 0.14 MeV \cite{Armstrong:1991yk}. 
Our estimate for the partial width of $X(3872)$ suggests that a 
new resonant $p \bar p$ experiment at GSI \cite{Ritman:2007ue}
or at Fermilab \cite{Kaplan:2007aj} should be able to measure the 
properties of the $X(3872)$ with comparable or higher accuracy 
than those of the $\chi_{c1}$.

It is also  useful to compare the production rates of $X(3872)$
and $\chi_{c1}$ in other processes.
The production rate of $X$ by the exclusive
decay $B^+ \to K^+ +X$ has been measured by the Belle and Babar
collaborations \cite{Choi:2003ue,Aubert:2004ns}.  
Dividing the measured product of the branching fractions for 
$B^+ \to K^+ +X$ and $X \to J/\psi \, \pi^+ \pi^-$
by the  branching fraction for $B^+ \to K^+ + \chi_{c1}$ 
\cite{Yao:2006px}, we obtain
\begin{equation}
\frac {\Gamma [B^+ \to K^+ + X]} {\Gamma [B^+ \to K^+ + \chi_{c1}]} =
\frac {0.023 \pm 0.005}{Br[X \to J/\psi \, \pi^+ \pi^-]} .
\label{RBK}
\end{equation}
Some information on the branching fraction for $X$ into 
$J/\psi \, \pi^+ \pi^-$ is provided by a measurement of the 
relative production rates for $J/\psi ~ \pi^+ \pi^-$ and 
$D^0 \bar D^0 \pi^0$ in the $X(3872)$ resonance region 
by the Belle Collaboration \cite{Gokhroo:2006bt}:
\begin{equation}
\frac{Br [X \to D^0 \bar D^0 \pi^0]}{Br[X \to J/\psi \, \pi^+ \pi^-]} 
= 8.8^{+3.1}_{-3.6} .
\label{RBrX}
\end{equation}
This measurement suggests that the branching fraction for $X$ into
$J/\psi ~ \pi^+ \pi^-$ is less than about 1/8.8, which implies 
that the production ratio in Eq.~(\ref{RBK}) is greater than 
about 0.2.

The production rates of $X(3872)$ and $\chi_{c1}$ are subject to 
various suppression factors. In the case of $X(3872)$, 
they include the factor $E_X^{1/2}$ associated with the large 
mean separation of its constituents. In the case of $\chi_{c1}$, 
they include a factor of $v^5$ associated with the small 
relative momentum $v$ of the $c$ and $\bar c$.  
These suppression factors apply equally well to all  
production processes. However there is one 
suppression factor in resonant $p \bar p$ production that does not 
apply to exclusive $B$ meson decay. 
This is the suppression factor associated with the
annihilation of all three quarks in the proton in the case of 
$\chi_{c1}$ and the annihilation of only two of the quarks in 
the case of $X(3872)$.  Because of this suppression factor, 
the production ratio 
for $X(3872)$ and $\chi_{c1}$ in resonant $p \bar p$ collisions 
should be larger than the production ratio in Eq.~(\ref{RBK}).
This is consistent with the estimated production ratio 
in resonant $p \bar p$ collisions obtained 
by dividing Eq.~(\ref{GamXppbar:est}) by the measured partial width 
for $\chi_{c1} \to p \bar p$.

In summary, our estimate of the partial width for 
$X(3872)$ into $p \bar p$ in Eq.~(\ref{GamXppbar:est})
indicates that it should be comparable to 
and perhaps even larger than that of the $\chi_{c1}$.
This implies that a future experiment on $p \bar p$ collisions 
at the $X(3872)$ resonance will allow the properties of this 
remarkable meson to be studied in great detail.

\begin{acknowledgments}
This research was supported in part by the Department of Energy
under grant DE-FG02-91-ER40690.
\end{acknowledgments}


\end{document}